\documentclass[a4paper,nofootinbib,notitlepage,longbibliography]{revtex4-1}
\pdfoutput=1

\usepackage{amsmath}
\usepackage{amssymb}
\usepackage{bbm}
\usepackage{booktabs}
\usepackage{braket}
\usepackage{dcolumn}
\usepackage{graphicx}
\usepackage{hyperref}
\usepackage[utf8]{inputenc}
\usepackage{mathtools}
\usepackage{multirow}
\usepackage{siunitx}
\usepackage[normalem]{ulem}
\usepackage{xcolor}
\usepackage{xspace}

\newcommand{\TeV}{\textrm{TeV}\xspace}

\newcommand{\EOS}{\texttt{EOS}\xspace}
\newcommand{\normflows}{\texttt{normflows}\xspace}
\newcommand{\pytorch}{\texttt{pytorch}\xspace}
\newcommand{\vecbeta}{\vec{\beta}\xspace}
\newcommand{\vecphi}{\vec{\varphi}\xspace}
\newcommand{\vecth}{\vec{\vartheta}\xspace}
\newcommand{\WET}{\text{WET}\xspace}
\newcommand{\tildeVub}{\ensuremath\tilde{V}_{ub}}
\newcommand{\wc}[2][\ensuremath{\ell}]{\ensuremath{\mathcal{C}_{#2}^{#1}}}
\newcommand{\op}[2][\ensuremath{\ell}]{\ensuremath{\mathcal{O}_{#2}^{#1}}}
\newcommand{\ublv}{\ensuremath{ub\ell\nu}\xspace}
\newcommand{\auc}{\text{AUC}\xspace}
\newcommand{\nf}{normalising flow\xspace}
\newcommand{\nfs}{normalising flows\xspace}
\newcommand{\smelli}{\texttt{smelli}\xspace}

\DeclarePairedDelimiter{\floor}{\lfloor}{\rfloor}
\DeclarePairedDelimiter{\ceil}{\lceil}{\rceil}

\makeatletter
\let\oldtheequation\theequation
\renewcommand\tagform@[1]{\maketag@@@{\ignorespaces#1\unskip\@@italiccorr}}
\renewcommand\theequation{(\oldtheequation)}
\makeatother

\begin{document}


\title{Testable Likelihoods for Beyond-the-Standard Model Fits}
\author{Anja Beck}
\email{anja.beck@cern.ch}
\affiliation{Department of Physics, University of Warwick, Coventry, CV4\,7AL, UK}
\author{Méril Reboud}
\email{merilreboud@gmail.com}
\affiliation{Institute for Particle Physics Phenomenology and Department of Physics, Durham University, Durham DH1 3LE, UK}
\author{Danny van Dyk}
\email{danny.van.dyk@gmail.com}
\affiliation{Institute for Particle Physics Phenomenology and Department of Physics, Durham University, Durham DH1 3LE, UK}

\begin{abstract}
Studying potential BSM effects at the precision frontier requires accurate transfer of
information from low-energy measurements to high-energy BSM models.
We propose to use \nfs to construct likelihood functions that achieve
this transfer. Likelihood functions constructed in this way provide the means to
generate additional samples and admit a ``trivial'' goodness-of-fit test in form of a $\chi^2$ test statistic.
Here, we study a particular form of \nf, apply it to a
multi-modal and non-Gaussian example, and quantify the accuracy of the likelihood function
and its test statistic.
\end{abstract}

\begin{flushright}
    IPPP/23/49
\end{flushright}
\vspace*{3\baselineskip}

\maketitle

\section{Introduction}
\label{sec:intro}

Contemporary experimental analyses at the Large Hadron Collider
have, so far, not been able to discover particles beyond the
Standard Model (BSM) at energy scales below $\simeq 1\,\TeV$.
As a consequence, model building has increasingly turned toward
using effective field theories (EFT) to describe any potential BSM
effects below these scales.
The Standard Model Effective Field Theory (SMEFT)~\cite{Buchmuller:1985jz} is one
of the main choices, as is the Higgs Effective Field Theory (HEFT)~\cite{Feruglio:1992wf}.
Constraining the EFT parameters is a challenge, due the large dimensionality
of the parameter space.
Taking the SMEFT as an example, the basis of operators
in the leading mass-dimension six Lagrangian
amounts to $2499$ independent EFT Wilson coefficients~\cite{Grzadkowski:2010es}.
The observed hierarchies among masses and mixing of quark flavours at low-energies
has inspired systematic approaches like Minimal Flavour Violation to reduce the number
of free parameters~\cite{DAmbrosio:2002vsn}.\\

Low-energy phenomena beside quark mixing have long been used to further our understanding
of BSM physics at large scales;
flavour-changing processes and anomalous electric dipole moments are excellent examples therof,
which contribute substantial statistical power to constrain BSM effects~\cite{Artuso:2022ouk}.
Flavour-changing processes in particular are commonly interpreted in a ``model-independent''
fashion by inferring their relevant parameters within another EFT, the Weak Effective Theory (WET)~\cite{Aebischer:2017gaw,Jenkins:2017jig,Jenkins:2017dyc}.
However, including low-energy constraints in this way provides for a challenge:
the interpretation of a majority of the low-energy constraints relies on our understanding
of hadronic physics in some capacity.
This leads to a proliferation of hadronic nuisance parameters,
which renders a global interpretation of all WET parameters impractical if not practically impossible.
Examples for this proliferation are plentiful in $b$-quark decay and include analyses of exclusive $b\to s\ell^+\ell^-$
processes with $113$ hadronic nuisance parameters and only $2$ parameters of interest~\cite{Gubernari:2022hxn}
as well as analyses of exclusive $b\to u\ell^-\bar\nu$ processes with $50$ hadronic nuisance parameters and only $5$ parameters
of interest~\cite{Leljak:2023gna}.
To overcome this problem, the following strategy has been devised~\cite{Aebischer:2017ugx}:
\begin{itemize}
    \item divide the free parameters of the WET into so-called ``sectors``, which are mutually independent to leading power in $G_F$;
    \item identify sets of observables that constrain a single sector of the WET and infer that sectors parameters;
    \item repeat for as many sectors as possible.
\end{itemize}
In this way, statistical constraints on the parameters in the individual WET sectors in form of posterior
densities can be obtained, providing a \emph{local picture} of the low-energy effects of BSM physics.

To gain a global picture of the BSM landscape, either in terms of the parameters of the SMEFT, the HEFT, or a UV-complete BSM model,
the previously obtained local constraints for individual WET sectors can, in principle, be used as a likelihood function.
On the physics side, this requires matching between the WET and the genuine parameters of interest, i.e, the parameters
of a UV-complete model, the SMEFT, or the HEFT.
In the case of the SMEFT, a complete matching of all dimension-6 operators to the full dimension-6 wet
has been achieved at the one-loop level~\cite{Dekens:2019ept}.

To date, the strategy outlined above has not yet been implemented: no library of statistical WET constraints is available.
The key obstacles in the implementation are not specific to the underlying physics.
Instead, they are the accurate transfer of the statistical results for even a single sector;
the potential to test a likelihood's goodness of fit through a suitable test statistic;
and the ease of use by providing a reference code that exemplifies the approach.\footnote{%
    An alternative to this strategy is to use the most constraining observables for each WET sector
    as low-energy likelihoods and use the latter directly within a global (SMEFT) likelihood.
    Despite the large number of nuisance parameters, this approach has been shown to be very useful
    and has been implemented as part of the \smelli software \cite{Aebischer:2018iyb}.
    However, it currently comes with the drawback of repeating the low-energy analyses
    for every point in the global (here: SMEFT) parameter space, leading to a waste of computational resources
    compared to the strategy proposed above. Improvements to \smelli overcoming this issue
    are presently under development.
}

Here, we present a small piece to solving the overall puzzle of how to
efficiently and accurately include the low-energy likelihoods in BSM fits: the construction of
a likelihood function that encodes the WET constraints and that admits a test statistic.
Our approach uses methods from the field of automatized learning and generational models.
We illustrate the approach at the hand of a concrete example likelihood, for which we use posterior samples for the WET
parameters obtained in the course of a previous analysis~\cite{Leljak:2023gna};
cf.~also a direct determination of SMEFT parameters in a similar setup~\cite{Greljo:2023bab}.
Our example likelihood is multi-modal, and the shape of each mode is distinctly non-Gaussian.
This renders the objective of providing a testable likelihood function in terms of the BSM parameters quite challenging.\\

The possible applications for obtaining such a likelihood function are two-fold:
first, it permits to generate additional samples that are (ideally) identically distributed
as the training samples.
Second, it provides a test statistic to use in a subsequent EFT fit.
We illustrate how \nfs make both applications possible,
by translating our example posterior ``target'' density to a unimodal multivariate Gaussian ``base'' density.
We propose a number of tests to check the quality of this translation, gauging the validity
of both applications.

\section{Preliminaries}
\label{sec:preliminaries}

\subsection{Notation and objective}
\label{sec:preliminaries:notation}

We begin by introducing our notation for the physics parameters
and their associated statistical quantities.
Let $P_T^*(\vecth \,|\, \WET)$ be the ``true'' posterior density for the low-energy \WET parameters $\vecth \in T \equiv \mathbb{R}^D$,
also known as the \WET Wilson coefficients.
We refer to the vector space $T$ as the \emph{target} space of dimension $D$.
In our later example, we will consider $T$ to be the space of \WET
Wilson coefficients in the $ub\ell\nu$ sector as studied in Ref.~\cite{Leljak:2023gna}.
We assume that we can access $P_T^*$ numerically by means of Monte Carlo importance samples.
Our intent is to construct a likelihood
\begin{equation}
    L(\vecphi) \equiv P_T(\vecth(\vecphi) \,|\, \WET)
\end{equation}
where $P_T$ is a model of $P_T^*$ and $\vecphi \in F$ represents some set of BSM physics parameters
within some vector space $F$.
In an envisaged SMEFT application, $F$ would represent the vector
space of SMEFT Wilson coefficients.
We stress at this point that, although $P_T^{(*)}$ is a probability density,
$L$ is in general not a probability density and is hence labelled a \emph{likelihood} function.\\

Next, we introduce a bijective mapping $f$ between our target space $T$
and some \textit{base} space $B$
\begin{equation}
    B \ni \vecbeta = f(\vecth)\,.
\end{equation}
Our objective is to find a mapping $f$ and its associated Jacobian $J_f \equiv \partial f/\partial \vecth$
\begin{equation}
    \vecbeta \sim P_B(\vecbeta) = J_f (P_T \circ f^{-1})(\vecbeta)
\end{equation}
that ensures that $P_B(\vecbeta)$ is multivariate standard Gaussian
density: $P_B(\vecbeta) = \mathcal{N}^{(D)}(\vecbeta | \vec{\mu} = \vec{0}, \Sigma = \mathbbm{1})$.
As a consequence, one finds immediately that the two-norm in base space
follows a $\chi^2$ distribution:
\begin{equation}
    \label{eq:teststatistic}
    ||\vecbeta||_2 \equiv ||f(\vecth)||_2
        \sim \chi^2(\nu = D)\,.
\end{equation}
The likelihood function is thus fully defined by the mapping $f$. A natural test statistic is provided
by the $\chi^2$ statistic for the squared $2$-norm in base space.

\subsection{A \nf for a real-valued non-volume preserving model}
\label{sec:preliminaries:nf}

The framework of \nfs~\cite{Tabak:2012} has been developed with the explicit intent to transform
an existing probability density to a (standard)normal density, i.e., to ``normalise`` the
density.
To this end, $f$ is constructed as a composition of $K$ individual bijective \emph{mapping layers} $f^{(k)}$,
\begin{equation}
    \label{eq:preliminaries:notation:nf}
    f = f^{(1)} \circ f^{(2)} \circ \dots \circ f^{(K)}\,.
\end{equation}
Here, we restrict ourselves to a particular class of mapping layer, following the
real non-volume preserving (RealNVP) model~\cite{Dinh2017:RealNVP}:
the affine coupling layers.
This class of layers are chosen here for their proven ability to map multi-modal
densities of real-valued parameters to a standard normal through non-volume-preserving
transformations~\cite{Dinh2017:RealNVP}.
Each of these layers has the structure
\begin{align}
    \label{eq:preliminaries:notation:individual_layer}
    f^{(k)}: \vec{x}\in \mathbb{R}^D
        & \mapsto f^{(k)}(\vec{x}) \in \mathbb{R}^D\\
    f^{(k)}(\vec{x})
        & = (p \circ a^{(k)})(\vec{x})\,.
\end{align}
In the above, $\vec{x} = (x_1, x_2, \dots, x_D)^T$, and $p$ is a permutation operation
\begin{align}
    p(\vec{x}) = (x_2, \dots, x_D, x_1)^T
\end{align}
with a trivial Jacobian $|J_{p}| = 1$.
The power of the RealNVP model lays in the use of \emph{affine coupling layers} $a^{(k)}$~\cite{Dinh2017:RealNVP}.
Each of these layers splits its input vector into two parts $\vec{x} = (\vec{x}_\text{lo}^T, \vec{x}_\text{hi}^T)^T$,
where $d_\text{lo} \equiv \dim \vec{x}_\text{lo} = \ceil{D/2}$ and $d_\text{hi} \equiv \dim \vec{x}_\text{hi} = \floor{D/2}$.
The affine coupling layers then map their inputs to
\begin{align}
    a^{(k)}(\vec{x}) = \left(\begin{matrix}
        \vec{x}_\text{lo}\\
        \vec{x}_\text{hi} \odot \exp(s^{(k)}(\vec{x}_\text{lo})) \oplus t^{(k)}(\vec{x}_\text{lo})
    \end{matrix}\right)\,,
\end{align}
where $s^{(k)}$ and $t^{(k)}$ are real-valued functions $\mathbb{R}^{d_\text{lo}} \mapsto \mathbb{R}^{d_\text{hi}}$,
and $\odot$ and $\oplus$ indicates element-wise multiplication and addition.
Since both $s^{(k)}$ and $t^{(k)}$ are independent of $\vec{x}_\text{hi}$, the Jacobian for
each affine coupling layer $a^{(k)}$ takes the simple form:
\begin{align}
    J_{a^{(k)}} \equiv \exp\left[ \sum_i s^{(k)}_i(\vec{x}_\text{lo}) \right]\,.
\end{align}
There are no further restrictions on the functions $s^{(k)}$ and $t^{(k)}$ and they are commonly
learned from data using a neural network~\cite{Stimper2023:normflows}.

\section[Physics Case: Semileptonic B decays]{\boldmath Physics Case: Semileptonic \texorpdfstring{$B$}{B} decays}
\label{sec:poc}

To illustrate the viability of our approach, we carry out a proof-of-concept (POC) study using
existing posterior samples~\cite{EOS-DATA-2023-01v2} from a previous BSM analysis~\cite{Leljak:2023gna}.
This analysis investigates the BSM reach in exclusive semileptonic $b\to u\ell\bar\nu$ processes within
the framework of the \ublv sector of the WET:
\begin{equation}
    \label{eq:poc:hamiltonian}
    \mathcal{H}^{ub\ell\nu}
        = -\frac{4 G_F}{\sqrt{2}} \tildeVub \sum_i \wc{i} \op{i} + \text{h.c.}\,.
\end{equation}
Here $\wc{i}$ represents a WET Wilson coefficient (i.e., a free parameter in the fit to data)
while $\op{i}$ represents a local dimension-six effective field operator (i.e., the source
of the $50$ hadronic nuisance parameters in the original analysis~\cite{Leljak:2023gna}).
The assumptions inherent to that analysis lead to a basis composed of five WET operators:
\begin{equation}
\begin{aligned}
    \op{V,L} & = \big[\bar{u} \gamma^\mu P_L  b\big]\, \big[\bar{\ell} \gamma_\mu      P_L \nu\big]\,, &
    \op{V,R} & = \big[\bar{u} \gamma^\mu P_R  b\big]\, \big[\bar{\ell} \gamma_\mu      P_L \nu\big]\,, \\
    \op{S,L} & = \big[\bar{u}            P_L  b\big]\, \big[\bar{\ell}                 P_L \nu\big]\,, &
    \op{S,R} & = \big[\bar{u}            P_R  b\big]\, \big[\bar{\ell}                 P_L \nu\big]\,, \\
    \op{T}   & = \big[\bar{u} \sigma^{\mu\nu} b\big]\, \big[\bar{\ell} \sigma_{\mu\nu} P_L \nu\big]\,.
\end{aligned}
\end{equation}

\begin{figure}[t]
    \centering
    \includegraphics[width=\textwidth]{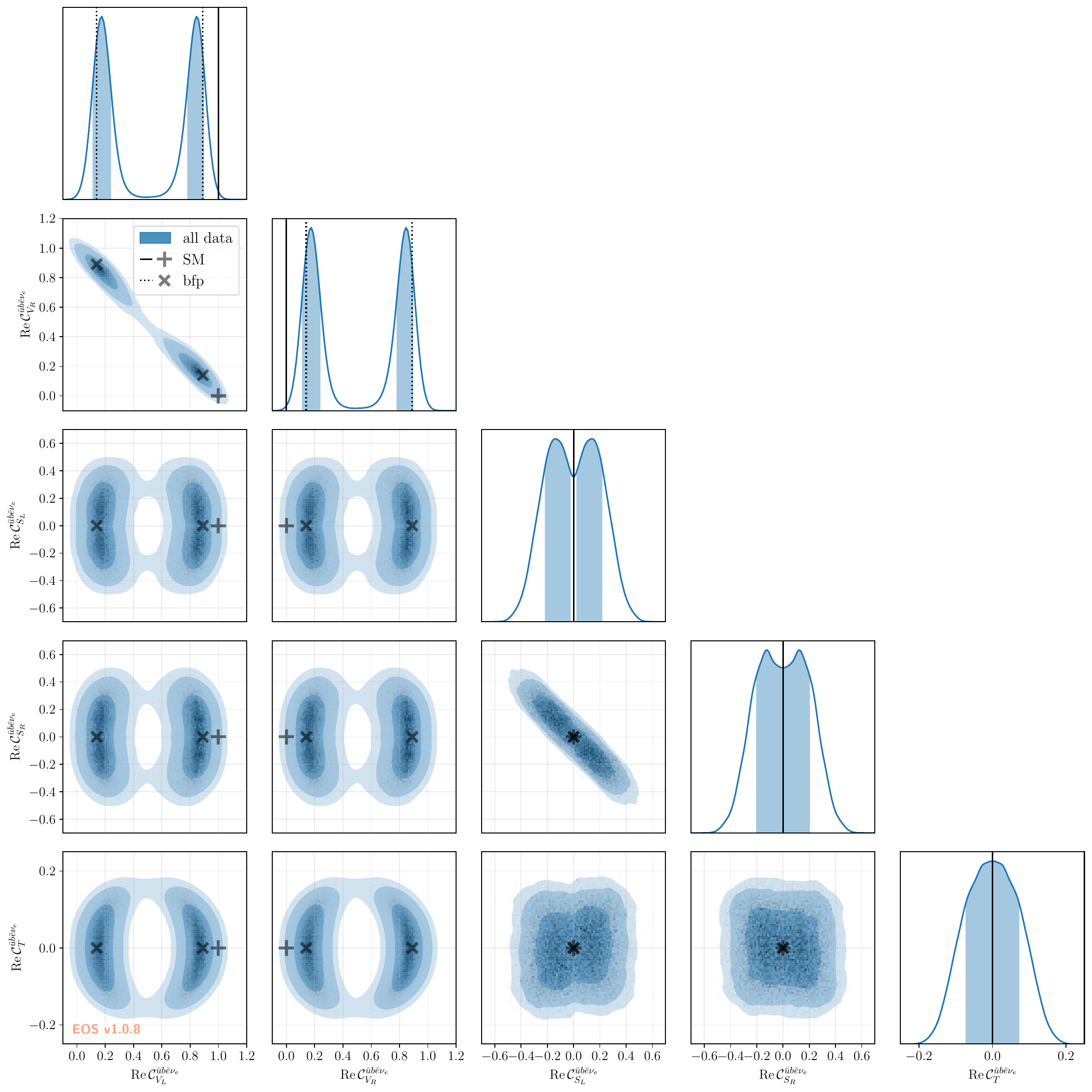}
    \caption{%
        Marginalized 5D Posterior for the parameters of interest $\vecth$ in the \ublv example.
        The black dots indicate the distribution of the posterior samples in form of a scatter plot.
        The blue-tinted areas correspond to the $68\%$, $95\%$, and $99\%$ probability regions as
        obtained from a smooth histogram based on kernel-density estimation.
    }
    \label{fig:poc:WET-posterior}
\end{figure}

For the purpose of a BSM study, only the marginal posterior of the WET parameters is of interest.
One readily obtains this marginal posterior by discarding the sample columns corresponding to any (hadronic) nuisance parameters.
A corner plot of this marginal posterior originally published in Ref.~\cite{Leljak:2023gna} is
shown in \autoref{fig:poc:WET-posterior}. As illustrated, the 5D marginal posterior is
very obviously non-Gaussian with a substantial number of isolated modes.
The samples obtained in this analysis are overlaid with a kernel-density estimate (KDE) of the marginalised posterior.
Although estimating the density is one of our main goals, the use of a KDE is neither sufficient nor advisable for the purpose we pursue.
First, KDEs are notorious for their computational costs.
Second, KDEs are sensitive to misestimation of densities close to the boundaries of the parameter space.
Last but not leas, KDEs do not provide a test statistic.

The objective of the POC study is now to determine whether a generative and testable likelihood can
be constructed from the available posterior samples by employing \nfs. To achieve this objective,
we investigate a two-dimensional subset of our posterior samples before we progress to the full five-dimensional problem.
For both cases, we use \nfs $f$ of the form shown in \autoref{eq:preliminaries:notation:nf},
i.e., a sequence of $K$ individual mapping layers \autoref{eq:preliminaries:notation:individual_layer},
each consisting of a composition of an affine coupling layer~\cite{Dinh2017:RealNVP} and a permutation.
This class of mapping layer is implemented as part of the \normflows software package~\cite{Stimper2023:normflows},
which we use to carry out our analysis. The \normflows software is based on
\pytorch~\cite{pytorch}, which is used in the training of the underlying neutral networks.

\begin{figure}[t]
    \centering
    \includegraphics[width=.5\textwidth]{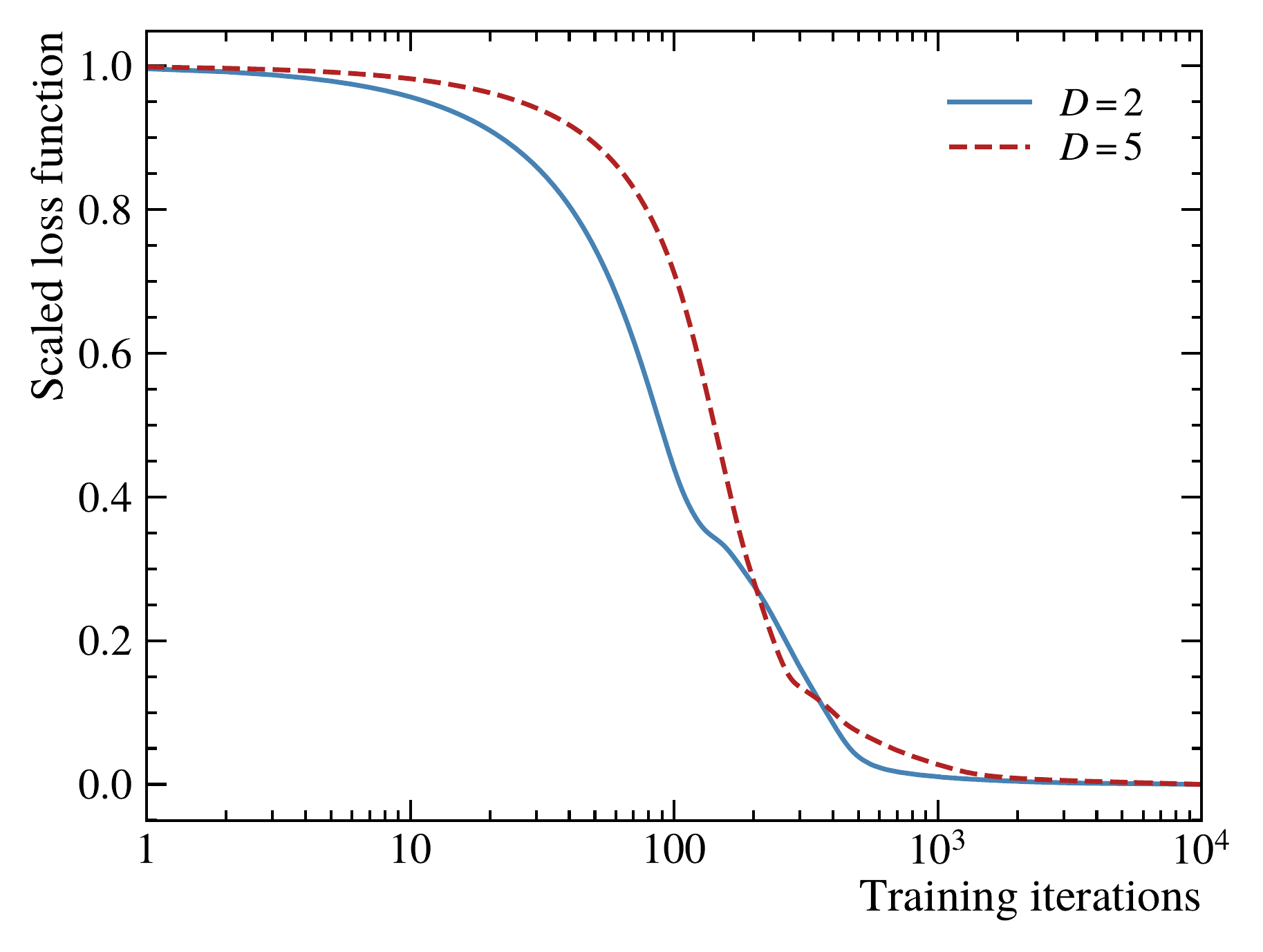}
    \caption{%
        Evolution of the loss function for the two- and five-dimensional case as the blue solid
        and red dashed lines, respectively.
        The loss values have been scaled so that all values are confined to the interval $[0, 1]$.
    }
    \label{fig:poc:2D:loss}
\end{figure}

We train the \nf on a total of $144k$ posterior samples using a total of $K=32$ mapping layers.
Each mapping layer is associated with a 4-layer perceptron, consisting of the input layer with
$\ceil{D/2}$ nodes, two hidden layers with $64$ nodes each, and the output layer with $2 \times \floor{D/2}$ nodes.
The outputs correspond to the vector-valued functions $s^{(k)}$ and $t^{(k)}$ as used in \autoref{eq:preliminaries:notation:individual_layer}.
We use the default choice of loss function provided by \normflows, the ``forward'' Kullback-Leibler divergence~\cite{papamakarios2021normalizing}
\begin{align}
    \mathrm{KL}(\vec{\alpha})
        & = - \mathbb{E}_{P_T^*(\vecth)}\left[\log P_T(\vecth \, | \, \vec{\alpha})\right] + \mathrm{const} \nonumber \\
        & = - \mathbb{E}_{P_T^*(\vecth)}\left[\log P_B(f(\vecth \, | \, \vec{\alpha})
            + \log \left|\det J_{f}(\vecth \, | \, \vec{\alpha}) \right|\right] + \mathrm{const}.
\end{align}
The loss function is minimised with respect to the neural network parameters $\vec{\alpha}$
using the ``Adam'' algorithm~\cite{kingma2017adam} as implemented in \pytorch.
Although we use a total of $10k$ optimisation iterations, we find that a close-to-optimal solution is found around the $2k$ iterations mark in both cases.
We show the evolution of the loss functions in terms of the number of optimisation steps in \autoref{fig:poc:2D:loss}.

\subsection{Application to a two-dimensional subset of samples}
\label{sec:2D_study}

In a first step, we perform a POC study at the hand of a subset of \WET Wilson coefficients.
The target space only consists of the two parameters:
\begin{equation}
    \label{eq:poc:2D:theta}
    \vecth = (\vartheta_1, \vartheta_2)^T \equiv (\text{Re}~\wc{V_L}, \text{Re}~\wc{S_L})^T \ .
\end{equation}
Our choice of 2D example exhibits two difficult features in target space as illustrated
in \autoref{fig:poc:WET-posterior}. First, the 2D marginal posterior is multi-modal.
Second, each mode is distinctly non-Gaussian and its isoprobability contours
resemble a bean-like shape. In the following discussion, we will frequently
refer to these shapes simply as ``beans''.
The central objective of this 2D POC study is to train a \nf $f$ such that the
base space variables
\begin{equation}
    \label{eq:poc:2D:beta}
    \vecbeta \equiv f(\vecth)\,.
\end{equation}
have a bivariate standard normal distribution: $\vecbeta \sim \mathcal{N}_2(\vec{0}, \mathbbm{1})$.\\

\begin{figure}[t]
    \centering
    \includegraphics[width=.35\textwidth]{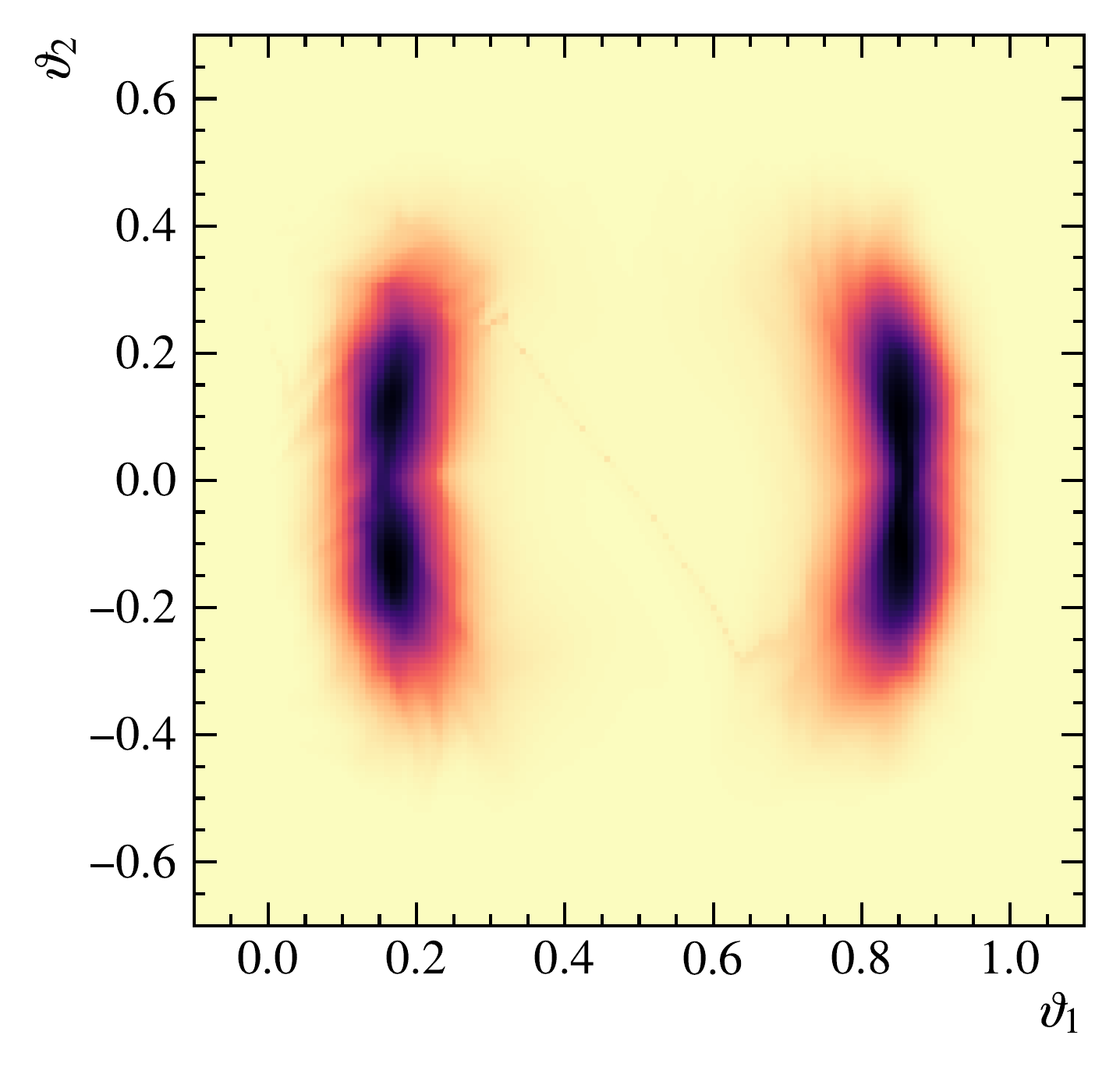}
    \includegraphics[width=.35\textwidth]{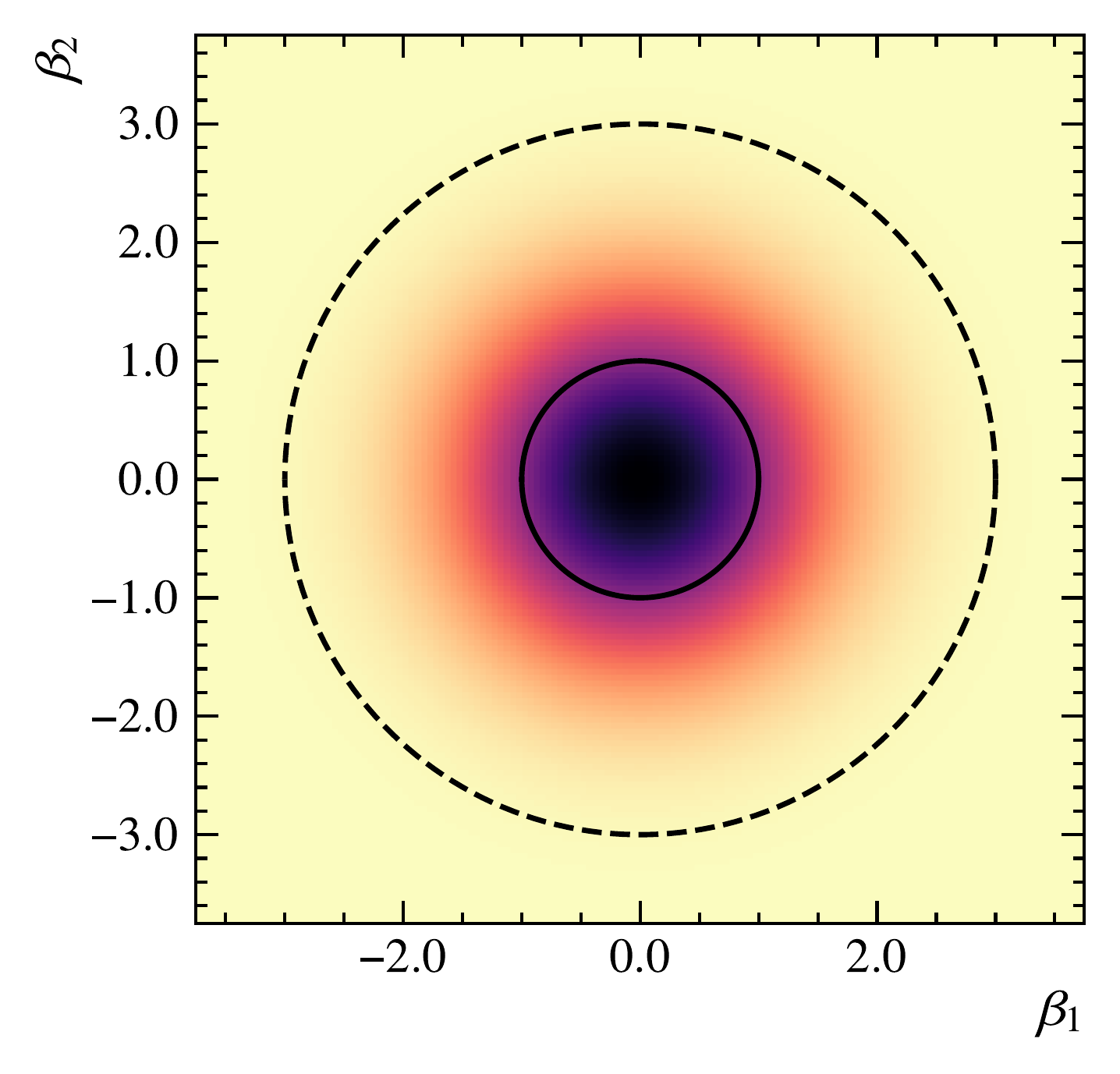}
    \includegraphics[width=.35\textwidth]{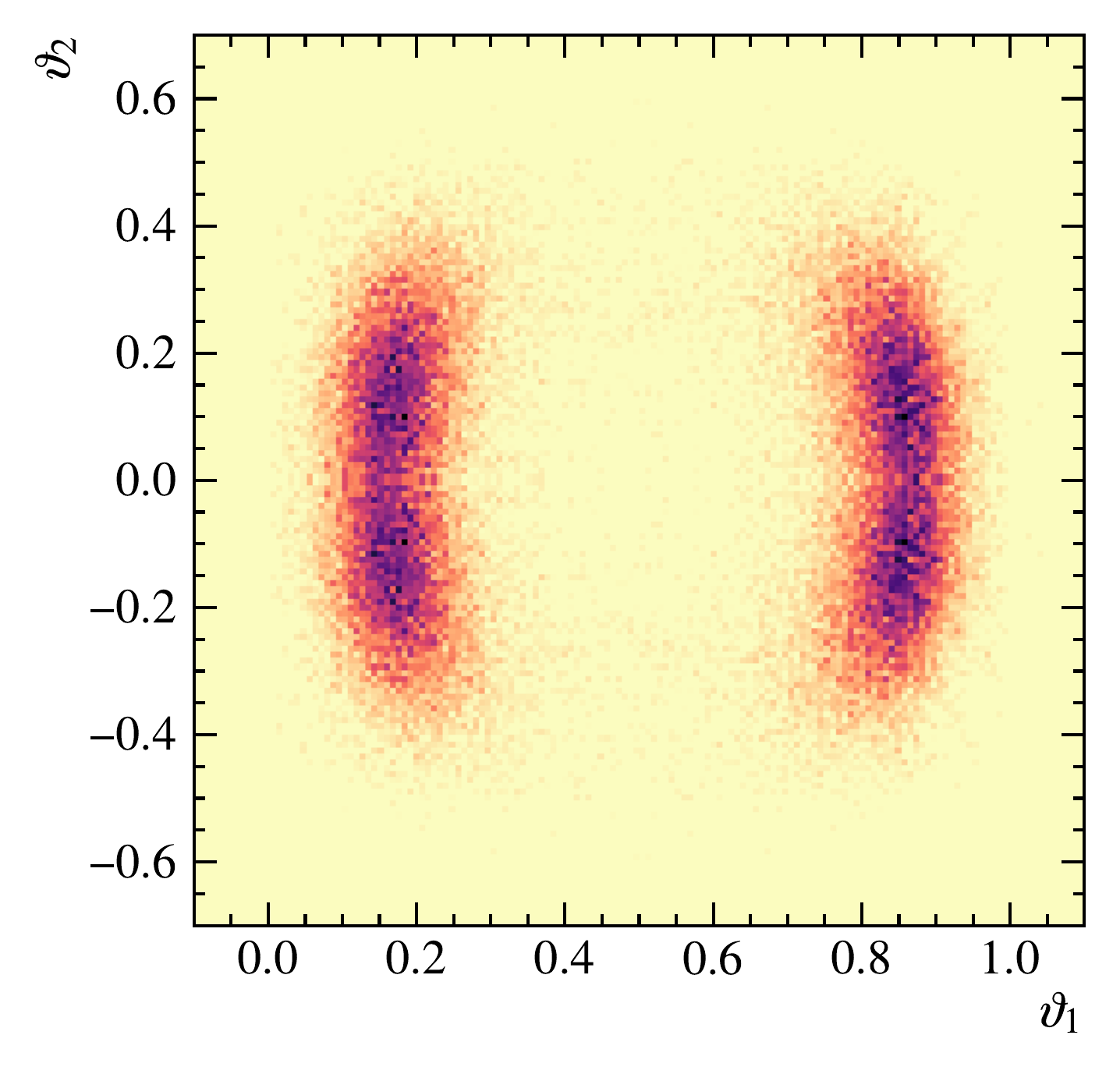}
    \includegraphics[width=.35\textwidth]{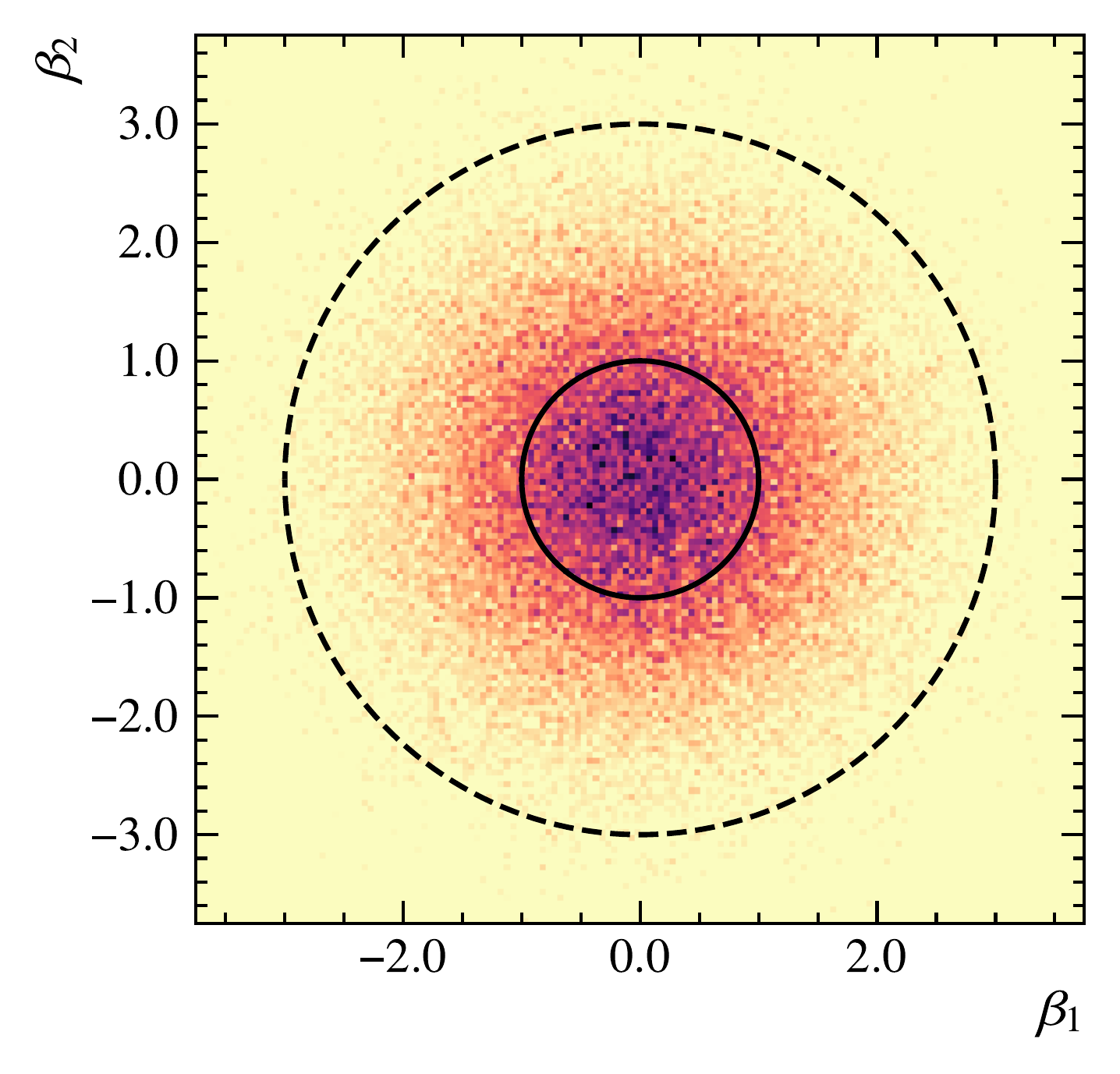}
    \caption{%
        Modelled (top) and empirical (bottom) distributions in the target (left) and base (right) space.
        Top: model in the target space after training (left) and base space (right).
        Bottom: distributions of the samples of the true posterior used for training the \nfs in the target space (left) and after transformation to the base space using the trained flow (right).
        The circles in the base space serve only to guide the eye and show the contours $||\vecbeta||_2=1$ and $||\vecbeta||_2=9$.
    }
    \label{fig:poc:2D:distributions}
\end{figure}

\begin{figure}[t]
    \centering
    \includegraphics[width=.35\textwidth]{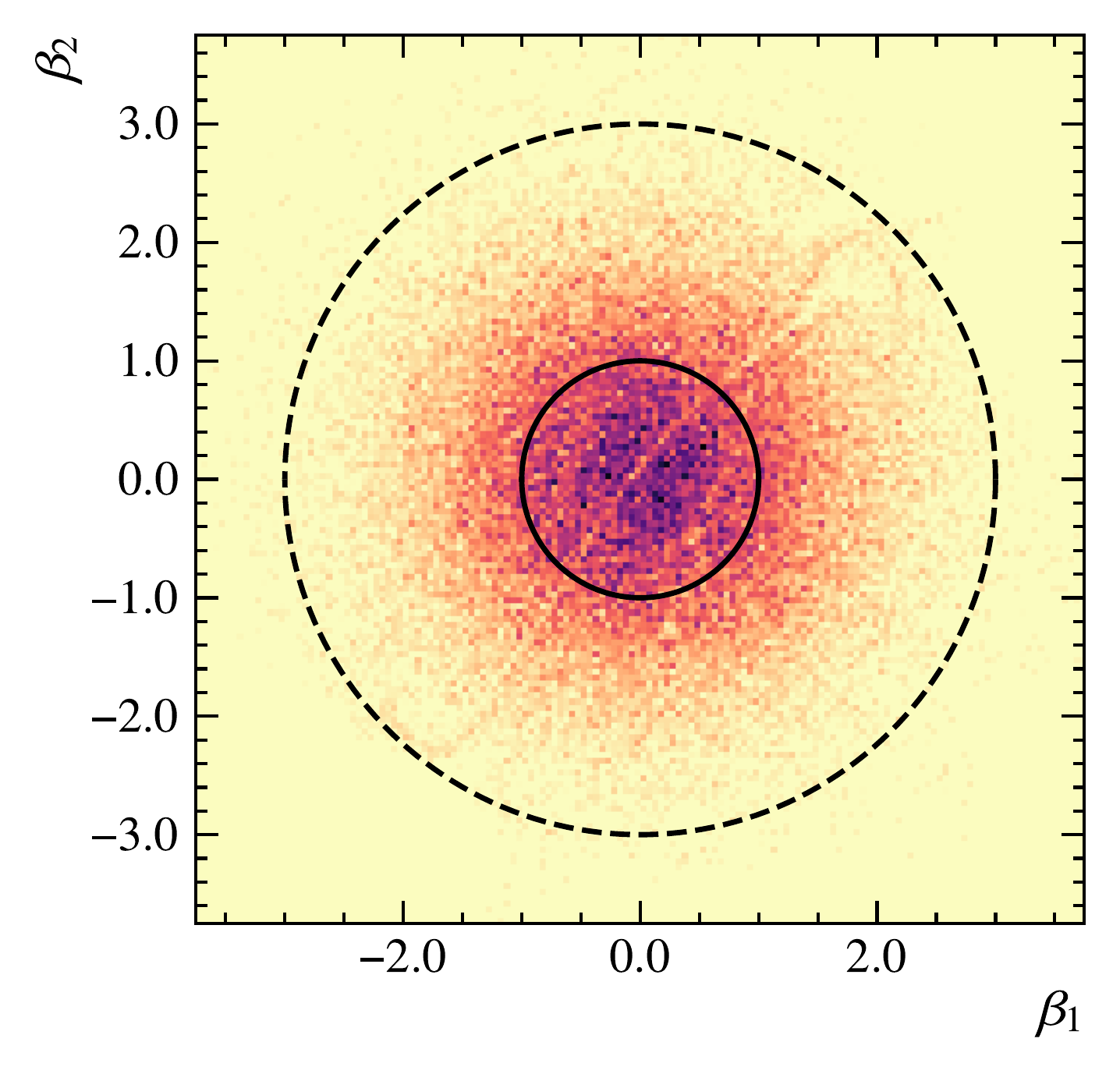}
    \includegraphics[width=.35\textwidth]{figures/data_gauss_2_10000.pdf}
    \caption{%
        Data distributions in the base space after 2000 (left) and 10000 (right) training iterations.
        The circles in the base space serve only to guide the eye and show the contours $||\vecbeta||_2=1$ and $||\vecbeta||_2=9$.
    }
    \label{fig:poc:2D:iterations}
\end{figure}

\paragraph{Distribution of the parameters in target and base space}~
In \autoref{fig:poc:2D:distributions} we show the result of the training in the target and base spaces.
The top plots contain the model in the target space after training and the base distribution used as starting point for the transformation.
The bottom plots show the distribution of the posterior samples in the target space, used in the training, and after transformation to the base space.
Visually, the trained model in the target space (top left) captures the main features of the training sample (bottom left).
Nevertheless, the model contains a faint filament connecting the left and right beans (top left), which is not present in the true posterior samples (bottom left).
Although this additional structure becomes less prominent as the value of the loss function decreases,
it never fully disappears using our analysis setup.
This additional feature translates to a less populated line across the sample distribution in the base space (bottom right).
This line however becomes invisible after a sufficient number of training iterations as is shown
in \autoref{fig:poc:2D:iterations}, where we juxtapose two transformation of the samples of $P_T^*$ from $T$ space to $B$
space using the \nfs trained with 2000 and trained with 10000 iterations.
We conclude that the presence of the filament requires careful diagnosis if the RealNVP model can be used to
achieve our stated objective: to obtain a generative and testable likelihood function.\\

\begin{figure}[t]
    \centering
    \includegraphics[width=.4\textwidth]{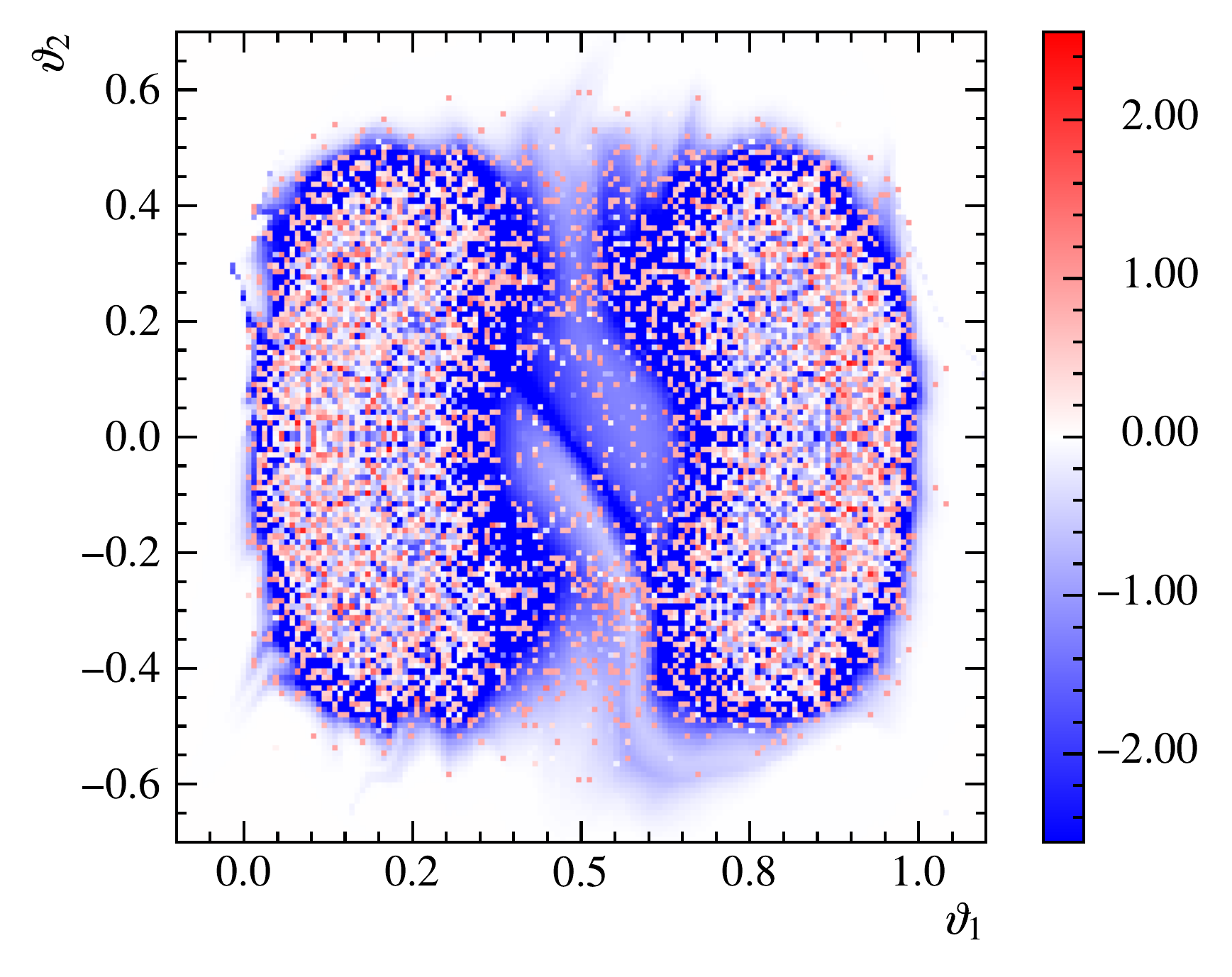}%
    \includegraphics[width=.4\textwidth]{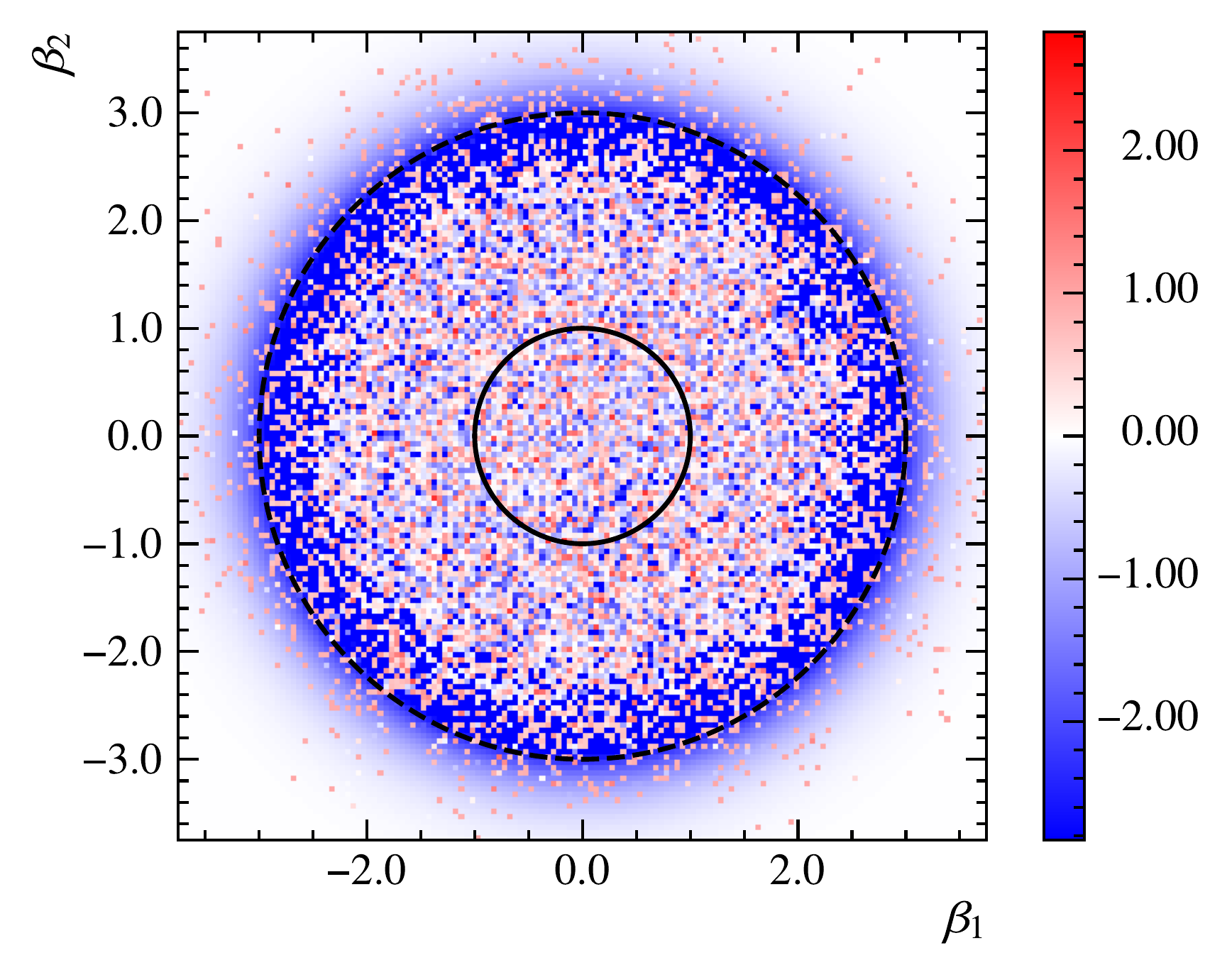}
    \caption{%
        Selected diagnostic plots for the proof of concept with the 2D subset of samples in the target (left) and base (right) space.
        Deviation of the model from the sample distribution in units of the uncertainty due to the finite size of the data sample.
        The circles in the base space serve only to guide the eye and show the contours $||\vecbeta||_2=1$ and $||\vecbeta||_2=9$.
    }
    \label{fig:poc:2D:diagnostics}
\end{figure}

\paragraph{Comparison of the modelled and true distributions}~
We investigate the quality of our trained \nfs with a comparison of the modelled and true distributions.
The comparison is first performed using histograms of the distributions in both the base and the target space.
The value $n_i$ of the true distribution in bin $x_i$ is obtained by dividing the number of samples in the bin by the total number of samples $N$.
The value of the modelled distribution is represented by the value of the modelled density at the bin centre, $P_X(x_i)$, multiplied with the area of the bin, $A_i \equiv A(x_i)$.
Our measure for the comparison is the deviation defined as
\begin{equation}
    \text{deviation} = \frac{n_i - P_X(x_i) A_i}{\sigma_i} \ ,
\end{equation}
where the uncertainty on the true distribution in bin $x_i$ is assumed to be
\begin{equation}
    \sigma_i = \frac{\sqrt{\max(1, N n_i)}}{N}
\end{equation}
The deviation is shown in \autoref{fig:poc:2D:diagnostics}.
As expected, the deviation is close to zero in highly populated regions and the connecting filament between the beans in the target space is clearly visible.
The larger deviations in sparsely populated regions are a result of the limited size of the data sample, where the red bins contain a single data point and the blue regions are empty bins.\\

\paragraph{Sample classification with a BDT}~
The next step is to quantify the quality of the \nfs in regard to the two primary purposes,
generating samples from the posterior and determining the test statistic for the actual distribution.
To this end, we train classifiers to distinguish samples generated from the true and modelled posterior distributions on the sample position in the base space, the position in target space, and the density in target space together with $\|\vecbeta\|_2$.
In all three cases, the classification is carried out with a Boosted Decision Tree (BDT)~\cite{Breiman}
with the default settings of the \texttt{XGBClassifier} class provided by the python library \texttt{xgboost}~\cite{Chen:2016:XST:2939672.2939785}.

The data generated from the true posterior is split into two equal parts, which are then only used for training and testing respectively.
A sample of the same statistical power is generated from the modelled distribution and also split into two equal parts.
The receiver-operator statistic is the dependence of the true positive rate on the false positive rate on the testing sample.
The area under its curve (\auc), for a set of features $x$ used in the classification, is a measure for the quality of the classification. Perfect classification results in $\auc(x) = 1$, while random classification results in $\auc(x) = 0.5$.\\

\paragraph{Differences between the modelled and the true posterior density}~
Training and testing the BDT on the position in target and base space, results in
\begin{equation}
    \auc(\vecth) = 0.577 \pm 0.002 \quad \text{and} \quad \auc(\vecbeta) = 0.549 \pm 0.003 \ ,
\end{equation}
respectively.
These low \auc values indicate that the true and modelled density are very similar.
However, they are significantly larger than 0.5 showing that the BDT can find exploitable differences.
The smaller \auc score in the base space compared to the target space is in line with the observation of a small connecting artefact in target space and the resulting diagonal gap in the base space, as discussed in relation to \autoref{fig:poc:2D:iterations} and \autoref{fig:poc:2D:diagnostics}.
As a sanity check, the BDT is also trained on two samples which are both generated from the model resulting in \auc values compatible with 0.5 as expected for two indistinguishable samples.\\

\begin{figure}[t]
    \centering
    \includegraphics[width=.5\textwidth]{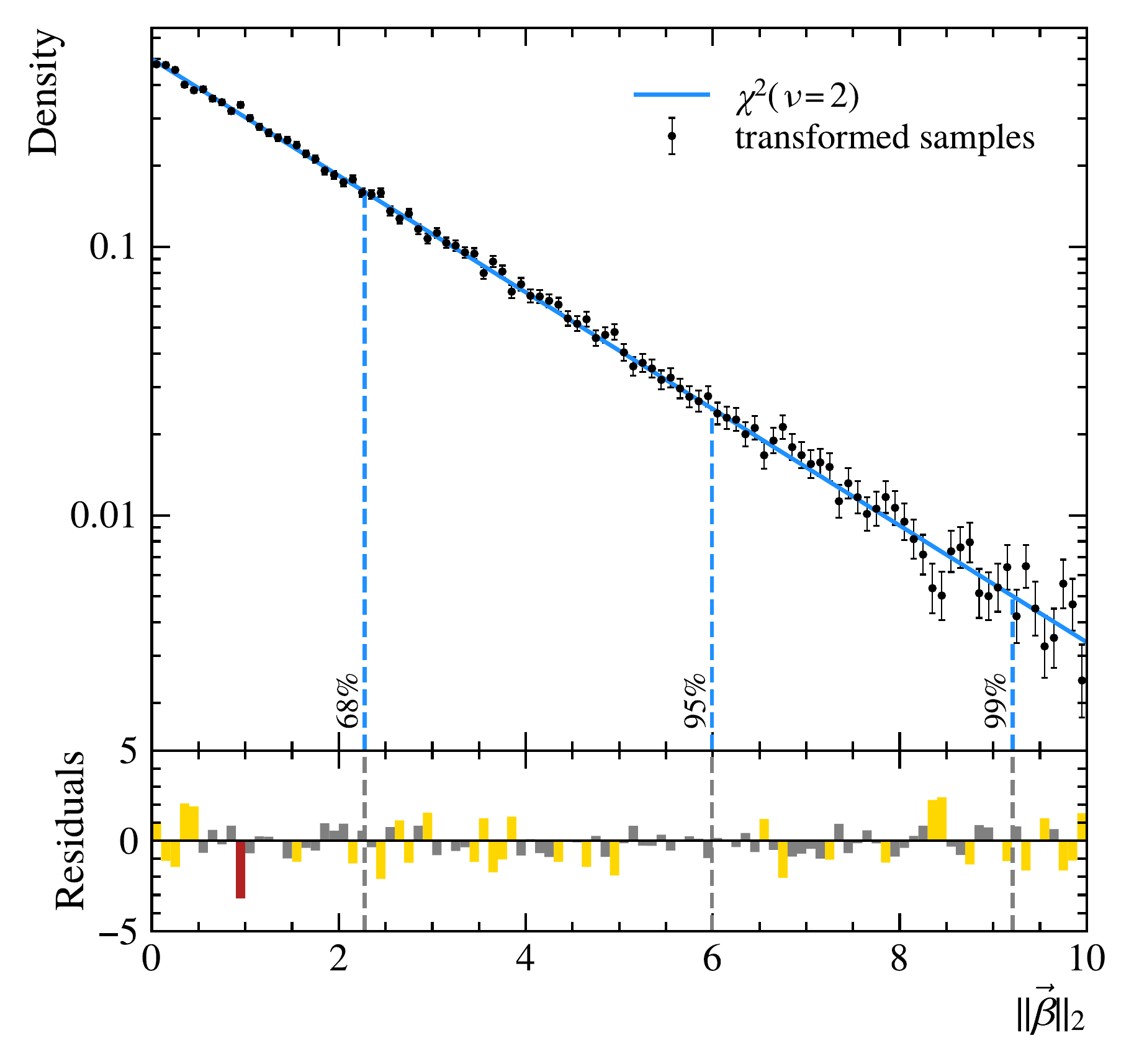}
    \caption{%
        Empirical distribution of $||\vecbeta||_2$ for the \ublv example using only two dimensions, see \autoref{eq:poc:2D:beta}.
        The colour of the bars representing the residuals, $r$, correspond to the magnitude, where grey, yellow, and red represent a residual with absolute value $|r|<1$, $1<|r|<3$, and $3<|r|$ respectively.
        The dashed vertical lines indicate the positions where the cumulative distribution function of $\chi^2(\nu=2)$ equals 68\%, 95\%, and 99\%.
    }
    \label{fig:poc:2D:empirical-chi2-dist}
\end{figure}

\paragraph{Testing the modelled test statistic}~
Finally, Fig.~\ref{fig:poc:2D:empirical-chi2-dist} shows the distribution of the 2-norm of the data samples in the base space.
They appear to follow a two-dimensional $\chi^2$ distribution as anticipated in Eq.~\ref{eq:teststatistic}.
We determine the reliability of the test statistic by investigating the ability of a BDT to distinguish the true and modelled posterior densities.
This BDT is trained on two variables: the posterior density in the target space and the 2-norm in the base space, $\|\vecbeta\|_2$, which resembles a $\chi^2$ distribution.

The posterior density is not available as a function.
We are therefore forced to use an empirical estimation.
Because the value of the density relies on the data itself, the \auc value for classification of two samples generated from the same model is larger than $0.5$ and depends on the estimation method used;
we investigate KDEs with different kernels as well as nearest neighbours density estimation.

The \auc scores for classification of samples generated from the true and modelled posterior density are always larger than the corresponding value for classification of two model-based samples.
The \auc value, obtained from estimating the density from the number of points with a distance below $0.015$ around a given point, is \mbox{$\auc(\text{density}) = 0.614\pm 0.006$} which is an increase of $0.039$ with respect to the baseline value calculated when classifying two model-based samples.\\

\paragraph{Conclusion of the two-dimensional study}
Our conclusion for this 2D POC study is that our approach is a viable one and
provides both an evaluatable likelihood and a reliable test statistic. Publishing 2D WET posteriors
in form of a \nf therefore clearly beats current best practices such as Gaussian approximations or Gaussian mixtures
models (as for example used in Ref.~\cite{Leljak:2023gna}) in terms of both usefulness and reliability.
We find that the presence of the filament does not adversely affect achieving our stated objectives
at the current level of precision. We leave an investigation as to other choices of mapping layers,
in particular autoregressive rational quadratic splines as used in Ref.~\cite{Reyes-Gonzalez:2023oei}, to future work.
Instead, we move on to apply our approach to the full five-dimensional POC.\\

\subsection{Application to the full five-dimensional samples}
We carry out the same analysis on the full five-dimensional posterior of \autoref{fig:poc:WET-posterior}.
This is a much more complicated case in terms of possible artefacts between beans.
Since this POC study is only illustrative, we do not try to optimise the training of the normalising flows and keep the same setup as for the two-dimensional case.
As visual comparisons of the distributions suffer from loss of information due to the projection from five to two dimension, we rely on the BDT tests that we introduced for the two-dimensional case.\\

\begin{figure}[t]
    \centering
    \includegraphics[width=.5\textwidth]{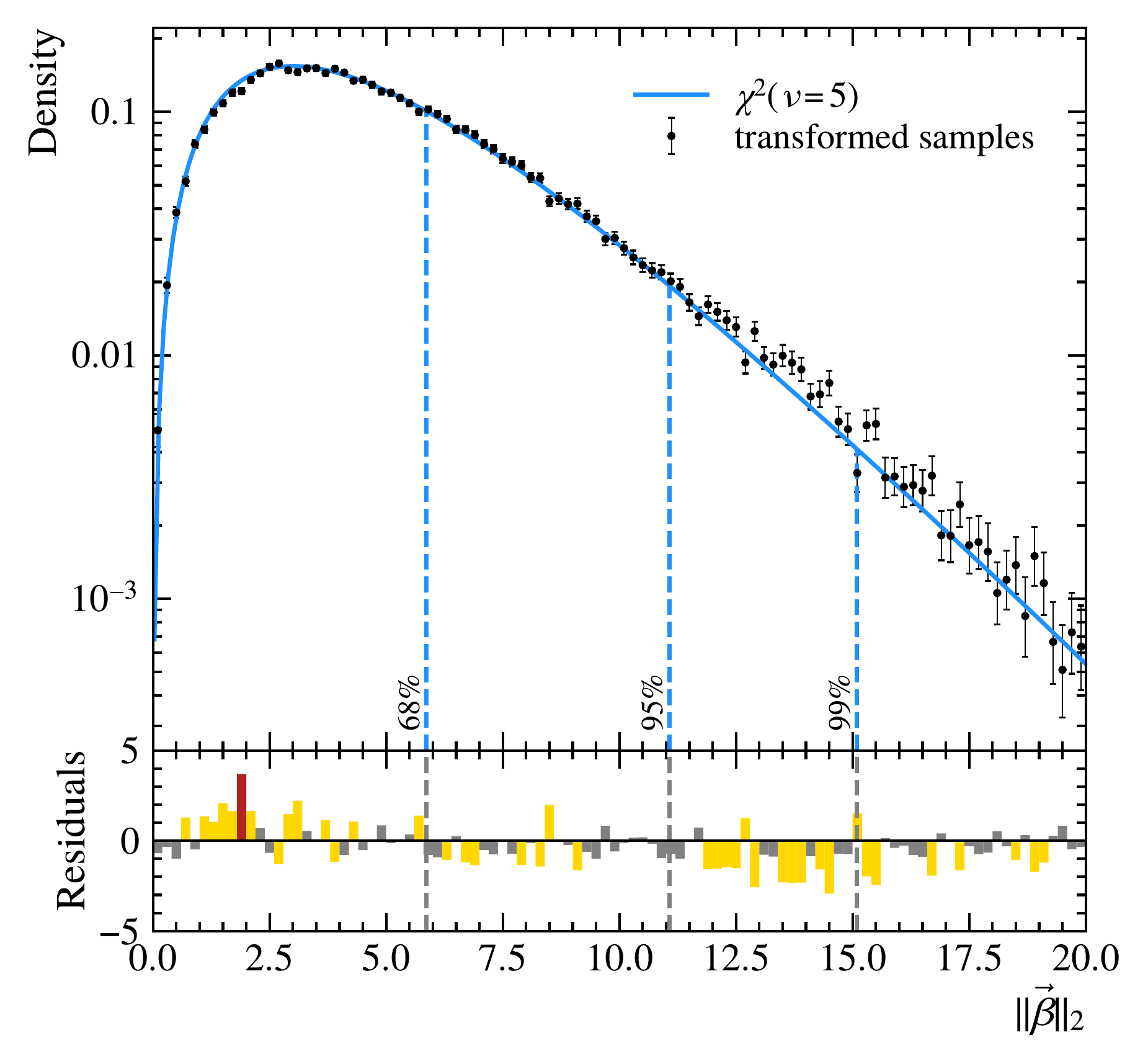}
    \caption{%
        Empirical distribution of $||\vecbeta||_2$ for the \ublv example using all five dimensions.
        The colour of the bars representing the residuals, $r$, correspond to the magnitude, where grey, yellow, and red represent a residual with absolute value $|r|<1$, $1<|r|<3$, and $3<|r|$ respectively.
        The dashed vertical lines indicate the positions where the cumulative distribution function of $\chi^2(\nu=5)$ equals 68\%, 95\%, and 99\%.
    }
    \label{fig:poc:5D:empirical-chi2-dist}
\end{figure}

Fig.~\ref{fig:poc:5D:empirical-chi2-dist} shows the distribution of the 2-norm of the samples transformed to the five-dimensional base space.
The distribution is overall consistent with a $\chi^2$ distribution with five dimensions.
the residuals reveal small systematic differences between the shape of the distribution of the data points and
the expected PDF. In particular, the residuals tend to aggregate positive values at small $\chi^2 < 5$
and negative values at large $10 < \chi^2 < 20$.
This conclusions are confirmed by the BDT classification in the base space which reaches an \auc score of
\begin{equation}
    \auc(\vecbeta) = 0.5708 \pm 0.0005 \ .
\end{equation}
This value is slightly larger than the value obtained in \autoref{sec:2D_study} but corroborates that the data sampled from the true posterior distribution resembles a five-dimensional Gaussian distribution after transformation with the trained \nf.\\

On the other hand, the closeness of the true and modelled distributions in the target space is much less convincing.
The BDT classification in the target space reaches \auc values close to one.
Similarly, training and testing the BDT on the value of the posterior density in target space and the 2-norm in base space gives $\auc(\text{density}) = 0.9150 \pm 0.0005$ which confirms a poor modelling of the true distribution.\\

We conclude that the full five-dimensional variable space presents a more complex challenge compared to the two-dimensional POC.
The basic \nf training as pursued in this work does not provide a satisfactory model for the multi-modal distribution in target space.
This is not surprising since we use the same setup for the two- and five-dimensional tests.
The small number of iterations needed to train the five-dimensional \nf, visible in \autoref{fig:poc:2D:loss},
indicates that more complicated models can be trained to achieve better performances.
The tuning of the model and the training procedure are however beyond the scope of this paper.

\section{Summary}
We show that \nfs can overcome two major hurdles to the full exploitation of 
statistical constraints on EFT parameters stemming from phenomenological analyses
of low-energy processes.
On the one hand, they enable sampling from a previous analysis' posterior distribution by converting it into a simple multivariate Gaussian density.
On the other hand, they provide a simple test statistics for the distribution.

We investigate this procedure by training a RealNVP \nf to a physics case that exhibits multi-modal posterior densities in two and five dimensions.
The compatibility of the modelled and the empirical posterior distributions are examined using different statistical tools.

Our conclusions are as follows.
When facing multi-modal distributions, the modelling of the true posterior densities
leads to the presence of artefacts as observed here. Nevertheless, these artefacts
do not prevent the successful usage of \nfs as proposed here.
Instead, they require careful (supervised) learning of the features of the true posterior density.
In the two-dimensional case, we have illustrated that both of our objectives, the sampling and the presence of a test statistics, are achieved.
However, applying our approach to the five-dimensional case without modification leads to less satisfactory performances.
Improvements to performance through detailed modelling, at the cost of more computationally-intensive training, are left for future work.

We are currently modifying the \EOS software~\cite{EOSAuthors:2021xpv} to
provide users with an interface that transparently makes
use of our proposed method of constructing low-energy
likelihoods for existing and future analyses of flavour-changing
processes.

\acknowledgments

We thank Eli Showalter-Loch and K. Keri Vos for their contribution to the early stages of the project.
D.v.D.~is grateful to Jack Araz, Sabine Kraml, Matthew Feickert, Daniel Maitre, and Humberto Reyes-Gonzalez for useful discussions.
D.v.D.~acknowledges support by the UK Science and Technology Facilities Council
(grant numbers ST/V003941/1 and ST/X003167/1).

\bibliography{references.bib}

\end{document}